\documentclass[aps,preprint,nofootinbib,superscriptaddress]{revtex4}%
\usepackage{hyperref}
\usepackage{amsmath}
\usepackage{amsfonts}
\usepackage{amssymb}
\usepackage{graphicx}%
\providecommand{\U}[1]{\protect\rule{.1in}{.1in}}

\begin{document}
\title{Extremal RN/CFT in Both Hands Revisited}
\author{En-Jui Kuo}
\email{eric1028.ep01@nctu.edu.tw}
\affiliation{Department of Electrophysics, National Chiao-Tung University, Hsinchu, Taiwan, R.O.C.}
\author{Yi Yang}
\email{yiyang@mail.nctu.edu.tw}
\affiliation{Department of Electrophysics, National Chiao-Tung University, Hsinchu, Taiwan, R.O.C.}
\date{\today }

\begin{abstract}
We study RN/CFT correspondence for four dimensional extremal
Reissner-Nordstrom black hole. We uplift the 4d RN black hole to a 5d rotating
black hole and make a geometric regularization of the 5d space-time. Both
hands central charges are obtained correctly at the same time by
Brown-Henneaux technique.

\end{abstract}
\maketitle

\section{Introduction}

To fully understand Bekenstein-Hawking entropy of a black hole in a
microscopic point of view is still a challenge. An important progress has been
made in \cite{0809.4266} by using the Brown-Henneaux technique \cite{BH}, i.e.
the Kerr/CFT correspondence. The central charge of the dual CFT reproduces the
exact Bekenstein-Hawking entropy of the 4-dimensional extremal Kerr black hole
by Cardy's formula. This method has been extensively studied for Kerr black
hole and other more general rotating black holes
\cite{0811.2225,0811.4393,0812.2918,0902.1001,0903.4030,1101.5136,1106.4148,1202.4156,1212.1959,1311.3384,1403.7352,1404.5260,1503.07861,1508.01583}%
. However, in the original Kerr/CFT method, the Virasoro algebra was realized
only from an enhancement of the rotational $U\left(  1\right)  $ isometry,
which corresponds to left hand central charge, not from the $SL\left(
2,R\right)  $. Later, it was found that the right hand central charge of
rotating black holes can be obtained by delicately choosing appropriate
boundary conditions,
\cite{0907.0303,0907.4272,0908.1121,1009.5039,1010.1379,1010.4291,1010.4549,1102.3423,1212.3031}%
.

Nevertheless, the method in Kerr/CFT correspondence cannot be directly applied
to non-rotating charge black holes, such as Reissner-Nordstrom (RN) black
hole. One found that the central charge for 4d RN black hole vanishes by
directly using the Brown-Henneaux technique. One possible way to solve the
problem is to uplift the space-time to higher dimensions. The left hand
central charge $c_{L}=6Q^{2}$\ of 4d extremal RN black hole was obtained by
uplifting it to 5 dimension
\cite{0811.4393,0902.4387,1001.2833,1010.1379,1106.4148,1106.4407}. The other
possible way is to consider a 2-dimensional effective theory of 4d extremal RN
black hole via a dimensional reduction following the idea in \cite{0908.1121},
by which the right hand central charge $c_{R}=6Q^{2}$ of 4d extremal RN way
obtained in \cite{0910.2076,1106.4407}. Although both hands central charges
can be obtained, but one has to use difference method. It is an interesting
question that whether we are able to calculate both hands central charges for
extremal RN black hole at the same time by choosing appropriate boundary
conditions as in the case of Kerr/CFT? The answer is yes, but with a geometric regularization.

In this paper, we calculate both hands central charges for 4d extremal RN
black hole by using the Brown-Henneaux technique. We first uplift the 4d RN
black hole to a 5d black hole, following the idea in \cite{1102.3423}, we then
deform the 5d black hole by a geometric regularization. Both hands central
charges $c_{R}=c_{L}=6Q^{2}$ are obtained in this way at the same time.

The paper is organized as follows. In section II, we briefly review the near
horizon geometry of 4d extremal RN black hole. We then uplift 4d RN black hole
to 5d and calculate both hands asymptotic Killing vectors in section III. In
section IV, we deform the 5d black hold by the geometric regularization to
obtain both hands central charges. Our conclusion and discussion are included
in section V.

\section{Near Horizon Geometry of 4d Extremal RN Black Hole}

In four dimension, the Einstein-Maxwell theory%
\begin{equation}
S_{4}=\frac{1}{16\pi G_{4}}\int d^{4}x\sqrt{-g_{4}}\left(  R_{4}-F^{2}\right)
,
\end{equation}
admits a unique spherical electro-vacuum solution, the RN black hole. The
metric of 4d non-extremal RN black hole is%
\begin{equation}
ds^{2}=-f\left(  r\right)  dt^{2}+\frac{dr^{2}}{f\left(  r\right)  }%
+r^{2}d\Omega_{2}^{2},
\end{equation}
where $d\Omega_{2}^{2}=d\theta^{2}+\sin^{2}\theta d\phi^{2}$ and%
\begin{equation}
f\left(  r\right)  =1-\dfrac{2m}{r}+\dfrac{q^{2}}{r^{2}}\text{, \ \ \ \ }%
A=-\frac{q}{r}dt.
\end{equation}
Two horizons are obtained by solving $f\left(  r\right)  =0$,%
\begin{equation}
r_{\pm}=m\pm\sqrt{m^{2}-q^{2}}.
\end{equation}
The Hawking temperature and black hole entropy can be calculated as follows,%
\begin{equation}
T_{H}=\frac{\kappa}{2\pi}=\dfrac{r_{+}-r_{-}}{4\pi r_{+}^{2}}\text{,
\ \ \ \ }S_{BH}=\frac{A}{4G_{4}}=\frac{\pi r_{+}^{2}}{G_{4}}.
\end{equation}
In the extremal limit of RN black hole, i.e. $r_{+}=r_{-}=m=q$, we have%
\begin{equation}
f\left(  r\right)  =\left(  1-\dfrac{q}{r}\right)  ^{2},
\end{equation}
and%
\begin{equation}
T_{H}=0\text{, \ \ \ \ }S_{BH}=\frac{\pi q^{2}}{G_{4}}=\pi Q^{2},
\label{BH entropy}%
\end{equation}
with we defined $Q\equiv q/\sqrt{G_{4}}$.

To consider the near horizon limit, $r\rightarrow r_{+}$, we make the
following coordinates transformation,%
\begin{equation}
\rho=\dfrac{r-r_{+}}{\epsilon q}\text{, \ \ \ \ }\tau=\dfrac{t}{q}\epsilon,
\end{equation}
and take the limit $\epsilon\rightarrow0$. Finally, the metric of 4d near
horizon extremal RN black hole is obtained as%
\begin{align}
ds^{2}  &  =q^{2}\left(  -\rho^{2}d\tau^{2}+\frac{d\rho^{2}}{\rho^{2}}%
+d\Omega_{2}^{2}\right)  ,\\
A  &  =q\rho d\tau.
\end{align}

\section{Central Charges by Uplifting to Five Dimension}

Direct calculation following the method of Kerr/CFT in \cite{0809.4266} shows
that the central charges of the 4d extremal RN black hole always
vanishes.\ This is because that the $U\left(  1\right)  $ symmetry of rotation
in Kerr black hole does not exist in the RN black hole, in which the $U\left(
1\right)  $ symmetry behaves as gauge symmetry. To get non-zero central
charge, a possible way is to uplift the 4d RN black hole into 5-dimensional
space-time with the extra dimension compactified on a circle as in
\cite{0811.4393,1106.4407}. The metric of the uplifted 5-dimensional black
hole can be expressed as,%

\begin{align}
ds^{2} &  =q^{2}\left(  -\rho^{2}d\tau^{2}+\frac{d\rho^{2}}{\rho^{2}}%
+d\Omega_{2}^{2}\right)  +\left(  dy+A\right)  ^{2}\nonumber\\
&  =\frac{q^{2}d\rho^{2}}{\rho^{2}}+q^{2}d\Omega_{2}^{2}+dy^{2}+2q\rho d\tau
dy,\label{5d metric}\\
A_{\left(  2\right)  } &  =\sqrt{3}q\rho d\tau\wedge dy.\label{2-form}%
\end{align}
where $A_{\left(  2\right)  }$ is a gauge 2-form and the new coordinate $y$ is
compactifed on a circle with a proper period \cite{1001.2833} $y=y+2\pi q$.
The metric (\ref{5d metric}) with the 2-form (\ref{2-form}) is a solution of
5d Einstein-Maxwell theory,%
\begin{equation}
S_{5}=\frac{1}{16\pi G_{5}}\int d^{4}x\sqrt{-g_{5}}\left(  R_{5}-\frac{1}%
{12}F_{\left(  3\right)  }^{2}\right)  ,
\end{equation}
where $G_{5}=2\pi qG_{4}$ and $F_{\left(  3\right)  }=dA_{\left(  2\right)  }%
$. We should notice that the 2-form field $A_{\left(  2\right)  }$ has no
contribution to the central charge based on the argument in
\cite{0902.1001,0902.4387}.

By choosing the following boundary condition,%
\begin{equation}
\left(
\begin{array}
[c]{lllll}%
h_{\tau\tau}=\mathcal{O}\left(  \rho^{2}\right)   & h_{\tau\rho}%
=\mathcal{O}\left(  \rho^{-2}\right)   & h_{\tau\theta}=\mathcal{O}\left(
\rho^{-1}\right)   & h_{\tau\phi}=\mathcal{O}\left(  \rho\right)   & h_{\tau
y}=\mathcal{O}\left(  1\right)  \\
& h_{\rho\rho}=\mathcal{O}\left(  \rho^{-3}\right)   & h_{\rho\theta
}=\mathcal{O}\left(  \rho^{-2}\right)   & h_{\rho\phi}=\mathcal{O}\left(
\rho^{-1}\right)   & h_{\rho y}=\mathcal{O}\left(  \rho^{-1}\right)  \\
&  & h_{\theta\theta}=\mathcal{O}\left(  \rho^{-1}\right)   & h_{\theta\phi
}=\mathcal{O}\left(  1\right)   & h_{\theta y}=\mathcal{O}\left(  1\right)  \\
&  &  & h_{\phi\phi}=\mathcal{O}\left(  \rho^{-1}\right)   & h_{\phi
y}=\mathcal{O}\left(  1\right)  \\
&  &  &  & h_{yy}=\mathcal{O}\left(  \rho\right)
\end{array}
\right)  ,\label{boundary condition}%
\end{equation}
the asymptotic Killing vector (AKV) is obtained as,%
\begin{equation}
\zeta=\epsilon\left(  \tau\right)  \partial_{\tau}+\left[  -\rho\partial
_{\tau}\epsilon\left(  \tau\right)  -\rho\partial_{y}\eta\left(  y\right)
\right]  \partial_{\rho}+\left[  \eta\left(  y\right)  -q\partial_{\tau}%
^{2}\epsilon\left(  \tau\right)  /\rho\right]  \partial_{y},
\end{equation}
where $\epsilon\left(  \tau\right)  $ and $\eta\left(  y\right)  $ are
arbitrary functions of $\tau$ and $y$, respectively. The left hand and right
hand AKVs reads%
\begin{subequations}
\begin{align}
\zeta^{L} &  =-\rho\partial_{y}\eta\left(  y\right)  \partial_{\rho}%
+\eta\left(  y\right)  \partial_{y},\\
\zeta^{R} &  =\epsilon\left(  \tau\right)  \partial_{\tau}-\rho\partial_{\tau
}\epsilon\left(  \tau\right)  \partial_{_{\rho}}-\frac{q}{\rho}\partial_{\tau
}^{2}\epsilon\left(  \tau\right)  \partial_{y}.
\end{align}
Since $y=y+2\pi q$ is periodic, we can express the function $\eta\left(
y\right)  $ as its Fourier bases $e^{-iny/q}$. Similarly, if we assign a
period for $\tau$\ with $\tau=\tau+2\pi\beta$ with $\beta$\ an arbitrary real
number, the function $\epsilon\left(  \tau\right)  $ can be expressed by
$e^{-in\tau/\beta}$. With appropriate normalizations, we can write%
\end{subequations}
\begin{equation}
\eta_{n}\left(  y\right)  =-qe^{-in\frac{y}{q}}\text{, \ \ \ \ }\epsilon
_{n}\left(  \tau\right)  =-\beta e^{-in\frac{\tau}{\beta}}.\label{modes}%
\end{equation}
The corresponding AKVs
\begin{subequations}
\label{AKV}%
\begin{align}
\zeta_{n}^{L} &  =-in\rho e^{-in\frac{y}{q}}\partial_{\rho}-qe^{-in\frac{y}%
{q}}\partial_{y},\\
\zeta_{n}^{R} &  =-\beta e^{-in\frac{\tau}{\beta}}\partial_{\tau}-in\rho
e^{-in\frac{\tau}{\beta}}\partial_{\rho}-\frac{n^{2}q}{\rho\beta}%
e^{-in\frac{\tau}{\beta}}\partial_{y},
\end{align}
composes two copies of Virasoro algebra without central extension as
expected,
\end{subequations}
\begin{subequations}
\label{Virasoro}%
\begin{align}
i\left[  \zeta_{m}^{L},\zeta_{n}^{L}\right]   &  =\left(  m-n\right)
\zeta_{m+n}^{L},\\
i\left[  \zeta_{m}^{R},\zeta_{n}^{R}\right]   &  =\left(  m-n\right)
\zeta_{m+n}^{R}.
\end{align}
To obtain the central charges, we define the asymptotic charge $Q_{\zeta}%
$\ associated to the AKV $\zeta$ as,%
\end{subequations}
\begin{equation}
Q_{\zeta}=\dfrac{1}{8\pi G_{5}}\int_{\partial\Sigma}k_{\zeta}\left[
h,g\right]  ,\label{asymptotic charge}%
\end{equation}
where the 2-form $k_{\zeta}$ is defined for a perturbation $h_{\mu\nu}%
$\ around the background metric $g_{\mu\nu}$ \cite{0811.2225},%
\begin{align}
k_{\zeta}\left[  h,g\right]   &  =\frac{1}{2}\left[  \zeta_{\nu}\nabla_{\mu
}h-\zeta_{\nu}\nabla_{\sigma}h_{\mu}^{\sigma}+\zeta_{\sigma}\nabla_{\nu}%
h_{\mu}^{\sigma}+\frac{1}{2}h\nabla_{\nu}\zeta_{\mu}-h_{\nu}^{\sigma}%
\nabla_{\sigma}\zeta_{\mu}\right.  \nonumber\\
&  \left.  +\frac{1}{2}h_{\nu\sigma}\left(  \nabla_{\mu}\zeta^{\sigma}%
+\nabla^{\sigma}\zeta_{\mu}\right)  \right]  \ast\left(  dx^{\mu}\wedge
dx^{\nu}\right)  ,\label{2 form}%
\end{align}
with $\ast$ denotes the Hodge dual, and $\Sigma$ is the 4-dimensional
equal-time hypersurface at $\tau=const$.

The Dirac bracket of the asymptotic charge is given by%
\begin{equation}
\left\{  Q_{\zeta_{m}},Q_{\zeta_{n}}\right\}  _{DB}=Q_{\left[  \zeta_{m}%
,\zeta_{n}\right]  }+\dfrac{1}{8\pi G_{5}}\int_{\partial\Sigma}k_{\zeta_{m}%
}\left[  L_{\zeta_{n}}\bar{g},\bar{g}\right]  .
\end{equation}
We define the generators $L_{n}$ of Virasoro algebra as%
\[
L_{n}\equiv Q_{\zeta_{n}}+\alpha\delta_{n},
\]
with $\alpha$ an arbitrary $c$-number. With the usual rule of Dirac brackets
$\left\{  Q_{\zeta_{m}},Q_{\zeta_{n}}\right\}  _{DB}\rightarrow-i\left[
Q_{\zeta_{m}},Q_{\zeta_{n}}\right]  $, the commutation relations of the
generators $L_{n}$ can be calculated as%
\begin{equation}
\left[  L_{m},L_{n}\right]  =i\left\{  Q_{\zeta_{m}},Q_{\zeta_{n}}\right\}
_{DB}=\left(  m-n\right)  L_{m+n}+\dfrac{i}{8\pi G_{5}}\int_{\partial\Sigma
}k_{\zeta_{m}}\left[  \mathcal{L}_{\zeta_{n}}\bar{g},\bar{g}\right]  ,
\end{equation}
where $\mathcal{L}$ is Lie derivative. Plugging AKVs (\ref{AKV}) into the
2-form (\ref{2 form}), we obtain%
\begin{subequations}
\begin{align}
\dfrac{i}{8\pi G_{5}}\int_{\partial\Sigma}k_{\zeta_{m}}^{L}\left[
\mathcal{L}_{\zeta_{n}}\bar{g},\bar{g}\right]   &  =\frac{q^{2}}{2G_{4}%
}\left(  m^{3}+m\right)  \delta_{m+n}=\frac{Q^{2}}{2}\left(  m^{3}+m\right)
\delta_{m+n},\\
\dfrac{i}{8\pi G_{5}}\int_{\partial\Sigma}k_{\zeta_{m}}^{R}\left[
\mathcal{L}_{\zeta_{n}}\bar{g},\bar{g}\right]   &  =0.
\end{align}
Comparing to the standard commutation relations of Virasoro algebra,%
\end{subequations}
\begin{equation}
\left[  L_{m},L_{n}\right]  =\left(  m-n\right)  L_{m+n}+\dfrac{c}{12}\left(
m^{3}-m\right)  \delta_{m+n},\label{standard}%
\end{equation}
the central charges are read off as\footnote{The term linear in $m$ plays no
essential role here, since it can be shifted arbitrarily by the $c$-number
$\alpha$ in the definition of the generators $L_{n}$.},%
\begin{equation}
c_{L}=6Q^{2}\text{, \ \ \ \ }c_{R}=0.
\end{equation}
We see that the above calculation only gives\ us the correct left hand central
charge with the right hand central charge vanishes. However, because there is
no gravitational anomaly, we expect that $c_{L}=c_{R}$. In next section, we
will make\ a geometric regularization of the 5d black hole and obtain the
correct both hands central charges together.

\section{Geometric Regularization}

At the boundary $\rho=const\rightarrow\infty$, the 5d metric (\ref{5d metric}%
)\ reduces to%
\begin{equation}
ds^{2}=\frac{q^{2}d\rho^{2}}{\rho^{2}}+q^{2}d\Omega_{2}^{2}+dy^{2}+2q\rho
d\tau dy\rightarrow2q\rho d\tau dy. \label{5d boundary}%
\end{equation}
In the above calculation, we took the "equal-time hypersurface" $\Sigma$ with
$\tau=const$ when we defined the asymptotic charge $Q_{\zeta}$ in
Eq.(\ref{asymptotic charge}). However, it is easy to see that $ds^{2}%
\rightarrow0$ for $\tau=const$ from Eq.(\ref{5d boundary}), so that $\Sigma$
is not a space-like surface but a light-like one on the boundary. Therefore
our naive choice of the "equal-time hypersurface" $\Sigma$ causes problem when
we consider the asymptotic behavior near the boundary. To define a space-like
surface at the boundary, we follows \cite{1102.3423} to regularize the 5d
geometry by a coordinates transformation,%

\begin{equation}
\tau^{\prime}=\tau-ay\text{, \ \ \ \ }y^{\prime}=y, \label{shift}%
\end{equation}
where $a$ is the regularization parameter. We should remark that the resulted
central charges should be independent of the regularization parameter $a$.

With the coordinates transformation (\ref{shift}), the 5d black hole metric
(\ref{5d metric}) becomes%
\begin{equation}
ds^{2}=\frac{q^{2}d\rho^{2}}{\rho^{2}}+q^{2}d\Omega_{2}^{2}+\left(
1+2aq\rho\right)  dy^{\prime2}+2q\rho d\tau^{\prime}dy^{\prime},
\end{equation}
which, at the boundary $\rho=const\rightarrow\infty$, reduces to%
\begin{equation}
ds^{2}\rightarrow2aq\rho dy^{\prime2}+2q\rho d\tau^{\prime}dy^{\prime}.
\end{equation}
Now the shifted "equal-time hypersurface" $\Sigma^{\prime}$ with $\tau
^{\prime}=const$ becomes a space-like surface $ds^{2}\rightarrow2aq\rho
dy^{\prime2}$.

We also note that, to respect the periodicity $y=y+2\pi q$,. we should fix the
period of $\tau$ by $\beta=aq$ in Eq.(\ref{modes}). and the AKVs can be
rewritten by the new coordinates $\left(  \tau^{\prime},y^{\prime}\right)  $
as%
\begin{subequations}
\begin{align}
\xi_{n}^{L} &  =aqe^{-in\frac{y^{\prime}}{q}}\partial_{\tau^{\prime}}-in\rho
e^{-in\frac{y^{\prime}}{q}}\partial_{\rho}-qe^{-in\frac{y^{\prime}}{q}%
}\partial_{y^{\prime}},\\
\xi_{n}^{R} &  =-\left(  aq+\frac{n^{2}}{\rho}\right)  e^{-in\left(
\frac{\tau^{\prime}}{a}+y^{\prime}\right)  /q}\partial_{\tau^{\prime}}-in\rho
e^{-in\left(  \frac{\tau^{\prime}}{a}+y^{\prime}\right)  /q}\partial_{\rho
}-\frac{n^{2}}{a\rho}e^{-in\left(  \frac{\tau^{\prime}}{a}+y^{\prime}\right)
/q}\partial_{y^{\prime}},
\end{align}
which also satisfy Virasoro algebra without central charge as in
Eq.(\ref{Virasoro}). Similar calculations can be carried on with the results,%
\end{subequations}
\begin{subequations}
\begin{align}
\dfrac{i}{8\pi G_{5}}\int_{\partial\Sigma}k_{\zeta_{m}}^{L}\left[
L_{\zeta_{n}}\bar{g},\bar{g}\right]   &  =\frac{q^{2}}{2G_{4}}\left(
m^{3}+m\right)  \delta_{m+n}=\frac{Q^{2}}{2}\left(  m^{3}+m\right)
\delta_{m+n},\\
\dfrac{i}{8\pi G_{5}}\int_{\partial\Sigma}k_{\zeta_{m}}^{R}\left[
L_{\zeta_{n}}\bar{g},\bar{g}\right]   &  =\frac{q^{2}}{2G_{4}}m^{3}%
\delta_{m+n}=\frac{Q^{2}}{2}m^{3}\delta_{m+n}.
\end{align}
Comparing to the standard commutation relations of Virasoro algebra
\ref{standard}, the above results lead to the correct both hands central
charges of 4d extremal RN black hole,%
\end{subequations}
\begin{equation}
c_{L}=c_{R}=6Q^{2}.\label{c}%
\end{equation}
We see that the central charges are indeed independent of the regularization
parameter $a$ as we promised.

Together with the CFT temperature of both hands \cite{0910.2076,1106.4407},%
\begin{equation}
T_{L}=\frac{1}{2\pi},\ \ \ \ T_{R}=0,
\end{equation}
we obtain the microscopic entropy by Cardy's formula from the central charges
(\ref{c}),%
\begin{equation}
S=\frac{\pi^{2}}{3}\left(  c_{L}T_{L}+c_{R}T_{R}\right)  =\pi Q^{2},
\end{equation}
which correctly reproduces the Bekenstein-Hawking entropy of the extremal RN
black hole (\ref{BH entropy}).

\section{Conclusion}

In this paper, we studied near horizon geometry of 4d extremal RN black hole.
We uplifted the 4d extremal RN black hole to a 5d black hole, and deformed the
5d black hole by a geometric regularization. Both hands central charges of 4d
extremal RN black hole $c_{L}=c_{R}=6Q^{2}$ were correctly obtained by using
the Brown-Henneaux technique. The crucial step to get the result is the
geometric regularization by which the shifted "equal-time hypersurface"
becomes space-like. The resulted central charges are independent of the
regularization as we expected.

\begin{acknowledgments}
This work is supported by the National Science Council (NSC
101-2112-M-009-005) and National Center for Theoretical Science, Taiwan.
\end{acknowledgments}

\bibliographystyle{unsrt}
\bibliography{Kerr-CFT}

\begin{thebibliography}{10}

\bibitem{0809.4266}
Monica Guica, Thomas Hartman, Wei Song, and Andrew Strominger.
\newblock {The kerr/cft correspondence}.
\newblock {\em arXiv preprint arXiv:0809.4266}, 2008.

\bibitem{BH}
J~David Brown and Marc Henneaux.
\newblock {Central charges in the canonical realization of asymptotic
  symmetries: an example from three dimensional gravity}.
\newblock {\em Communications in Mathematical Physics}, 104(2):207--226, 1986.

\bibitem{0811.2225}
H~L{\"u}, Jianwei Mei, and CN~Pope.
\newblock {Kerr-AdS/CFT correspondence in diverse dimensions}.
\newblock {\em Journal of High Energy Physics}, 2009(04):054, 2009.

\bibitem{0811.4393}
Thomas Hartman, Keiju Murata, Tatsuma Nishioka, and Andrew Strominger.
\newblock {CFT duals for extreme black holes}.
\newblock {\em Journal of High Energy Physics}, 2009(04):019, 2009.

\bibitem{0812.2918}
David~DK Chow, Mirjam Cveti{\v{c}}, H~L{\"u}, and CN~Pope.
\newblock {Extremal black hole/CFT correspondence in (gauged) supergravities}.
\newblock {\em Physical Review D}, 79(8):084018, 2009.

\bibitem{0902.1001}
Geoffrey Compere, Keiju Murata, and Tatsuma Nishioka.
\newblock Central charges in extreme black hole/cft correspondence.
\newblock {\em Journal of High Energy Physics}, 2009(05):077, 2009.

\bibitem{0903.4030}
Wen-Yu Wen.
\newblock {Holographic descriptions of (near-) extremal black holes in five
  dimensional minimal supergravity}.
\newblock {\em arXiv preprint arXiv:0903.4030}, 2009.

\bibitem{1101.5136}
Steven Carlip.
\newblock {Extremal and nonextremal Kerr/CFT correspondences}.
\newblock {\em Journal of High Energy Physics}, 2011(4):1--17, 2011.

\bibitem{1106.4148}
Bin Chen and Jia-ju Zhang.
\newblock {Novel CFT duals for extreme black holes}.
\newblock {\em Nuclear Physics B}, 856(2):449--474, 2012.

\bibitem{1202.4156}
Jianwei Mei.
\newblock {On the general Kerr/CFT correspondence in arbitrary dimensions}.
\newblock {\em Journal of High Energy Physics}, 2012(4):1--18, 2012.

\bibitem{1212.1959}
Bin Chen and Jia-ju Zhang.
\newblock {RN/CFT correspondence from thermodynamics}.
\newblock {\em Journal of High Energy Physics}, 2013(1):1--30, 2013.

\bibitem{1311.3384}
Cheng-Yong Zhang, Yu~Tian, and Xiao-Ning Wu.
\newblock {Generalized Kerr/CFT correspondence with electromagnetic field}.
\newblock {\em Classical and Quantum Gravity}, 31(8):085009, 2014.

\bibitem{1403.7352}
Cheng-Yong Zhang, Yu~Tian, Xiao-Ning Wu, and Shao-Jun Zhang.
\newblock {Entropy of the isolated horizon from the surface term of
  gravitational action}.
\newblock {\em Classical and Quantum Gravity}, 31(19):195002, 2014.

\bibitem{1404.5260}
Avirup Ghosh.
\newblock {Note on Kerr/CFT correspondence in a first order formalism}.
\newblock {\em Physical Review D}, 89(12):124035, 2014.

\bibitem{1503.07861}
G~Comp{\`e}re, K~Hajian, A~Seraj, and MM~Sheikh-Jabbari.
\newblock {Extremal Rotating Black Holes in the Near-Horizon Limit: Phase Space
  and Symmetry Algebra}.
\newblock {\em arXiv preprint arXiv:1503.07861}, 2015.

\bibitem{1508.01583}
Marco Astorino.
\newblock {Magnetised Kerr/CFT correspondence}.
\newblock {\em Physics Letters B}, 751:96--106, 2015.

\bibitem{0907.0303}
Yoshinori Matsuo, Takuya Tsukioka, and Chul-Moon Yoo.
\newblock {Another realization of Kerr/CFT correspondence}.
\newblock {\em Nuclear physics B}, 825(1):231--241, 2010.

\bibitem{0907.4272}
Yoshinori Matsuo, Takuya Tsukioka, and Chul-Moon Yoo.
\newblock {Yet another realization of Kerr/CFT correspondence}.
\newblock {\em EPL (Europhysics Letters)}, 89(6):60001, 2010.

\bibitem{0908.1121}
Alejandra Castro and Finn Larsen.
\newblock {Near extremal Kerr entropy from AdS2 quantum gravity}.
\newblock {\em Journal of High Energy Physics}, 2009(12):037, 2009.

\bibitem{1009.5039}
Monica Guica and Andrew Strominger.
\newblock {Microscopic realization of the Kerr/CFT correspondence}.
\newblock {\em Journal of High Energy Physics}, 2011(2):1--20, 2011.

\bibitem{1010.1379}
Bin Chen, Chiang-Mei Chen, and Bo~Ning.
\newblock {Holographic Q-picture of Kerr--Newman--AdS--dS black hole}.
\newblock {\em Nuclear Physics B}, 853(1):196--209, 2011.

\bibitem{1010.4291}
Tatsuo Azeyanagi, Noriaki Ogawa, and Seiji Terashima.
\newblock {Emergent AdS3 in the zero entropy extremal black holes}.
\newblock {\em Journal of High Energy Physics}, 2011(3):1--27, 2011.

\bibitem{1010.4549}
Yoshinori Matsuo and Tatsuma Nishioka.
\newblock {New near horizon limit in Kerr/CFT}.
\newblock {\em Journal of High Energy Physics}, 2010(12):1--24, 2010.

\bibitem{1102.3423}
Tatsuo Azeyanagi, Noriaki Ogawa, and Seiji Terashima.
\newblock {On non-chiral extension of Kerr/CFT}.
\newblock {\em Journal of High Energy Physics}, 2011(6):1--25, 2011.

\bibitem{1212.3031}
Ee~Chang-Young and Myungseok Eune.
\newblock {Nonextremal Kerr/CFT on a stretched horizon}.
\newblock {\em arXiv preprint arXiv:1212.3031}, 2012.

\bibitem{0902.4387}
Ahmad Ghodsi and Mohammad~R Garousi.
\newblock {The RN/CFT correspondence}.
\newblock {\em Physics Letters B}, 687(1):79--83, 2010.

\bibitem{1001.2833}
Chiang-Mei Chen, Ying-Ming Huang, and Shou-Jyun Zou.
\newblock {Holographic duals of near-extremal Reissner-Nordstr{\o}m black
  holes}.
\newblock {\em Journal of High Energy Physics}, 2010(3):1--14, 2010.

\bibitem{1106.4407}
Chiang-Mei Chen and Jia-Rui Sun.
\newblock {Holographic Dual of the Reissner-Nordstr{\"o}m Black Hole}.
\newblock In {\em Journal of Physics: Conference Series}, volume 330, page
  012009. IOP Publishing, 2011.

\bibitem{0910.2076}
Chiang-Mei Chen, Jia-Rui Sun, and Shou-Jyun Zou.
\newblock {The RN/CFT correspondence revisited}.
\newblock {\em Journal of High Energy Physics}, 2010(1):1--13, 2010.

\bibitem{1004.3963}
Chiang-Mei Chen and Jia-Rui Sun.
\newblock {Hidden conformal symmetry of the Reissner-Nordstr{\"o}m black
  holes}.
\newblock {\em Journal of High Energy Physics}, 2010(8):1--11, 2010.

\bibitem{1006.4092}
Chiang-Mei Chen, Ying-Ming Huang, Jia-Rui Sun, Ming-Fan Wu, and Shou-Jyun Zou.
\newblock Holographic dual of the dyonic reissner-nordstr{\"o}m black hole.
\newblock {\em Physical Review D}, 82(6):066003, 2010.

\end{thebibliography}

\end{document}